\documentclass[sigconf]{arxiv_cls}

\usepackage{balance}  % do not change this line -- unless you manually balance your last page

\usepackage{booktabs}
\usepackage{subfig}

\usepackage{algorithm}
\usepackage{algorithmic}

\usepackage{tikz}
\usetikzlibrary{decorations.pathreplacing,calc,tikzmark}

\usepackage{eqparbox}

\usepackage{balance}
\begin{document}

\title{Microscopic Traffic Simulation by\\ Cooperative Multi-agent Deep Reinforcement Learning}

% replace this with your author block!
\author{Giulio Bacchiani}
%\authornote{}
\orcid{0000-0003-3454-4287}
\affiliation{%
  \institution{VisLab - University of Parma}
  %\streetaddress{Parco Area delle Scienze, Pad. 33}
  \city{Parma} 
  \state{Italy} 
  \postcode{43124}
}
\email{giulio.bacchiani@studenti.unipr.it}

\author{Daniele Molinari}
\affiliation{%
  \institution{VisLab}
  \city{Parma} 
  \state{Italy} 
  \postcode{43124}
}
\email{dmolina@vislab.it}

\author{Marco Patander}
\affiliation{%
	\institution{VisLab}
	\city{Parma} 
	\state{Italy} 
	\postcode{43124}
}
\email{patander@vislab.it}

\begin{abstract}  % put your abstract here!
Expert human drivers perform actions relying on traffic laws and their previous experience.
While traffic laws are easily embedded into an artificial brain, modeling human complex behaviors which come from past experience is a more challenging task.
One of these behaviors is the capability of communicating intentions and negotiating the right of way through driving actions, as when a driver is entering a crowded roundabout and observes other cars movements to guess the best time to merge in.
In addition, each driver has its own unique driving style, which is conditioned by both its personal characteristics, such as age and quality of sight, and external factors, such as being late or in a bad mood. 
For these reasons, the interaction between different drivers is not trivial to simulate in a realistic manner.
In this paper, this problem is addressed by developing a microscopic simulator using a \textit{Deep Reinforcement Learning Algorithm} based on a combination of visual frames, representing the perception around the vehicle, and a vector of numerical parameters.
In particular, the algorithm called \textit{Asynchronous Advantage Actor-Critic} has been extended to a multi-agent scenario in which every agent needs to learn to interact with other similar agents. 
Moreover, the model includes a novel architecture such that the driving style of each vehicle is adjustable by tuning some of its input parameters, permitting to simulate drivers with different levels of aggressiveness and desired cruising speeds.
\end{abstract}

 \maketitle

%%%%%%%%%%%%%%%%%%%%%%%%%%%%%%%%%%%%%%%%%%%%%%%%%%%%%%%%%%%%%%%%%%%%%%%%%%%%%%%%%%%%%%%%%%%%%%%%%%%%%%%%%
%% start of main body of paper

%\input{chapters/abstract}
%\vspace*{2.0em}
\section{Introduction}
The development of autonomous vehicles is a topic of great interest in recent years, gaining more and more attention both from academy and industry.
An interesting challenge is teaching the autonomous car to interact and thus implicitly communicate with human drivers about the execution of particular maneuvers, such as entering a roundabout or an intersection. This is a mandatory request since the introduction of self-driving cars onto public roads is going to be gradual, hence human- and self-driving vehicles have to cohabit the same spaces.
In addition, it would be desirable that this communication would resembles the one which takes place every day on the streets between human beings, so that human drivers does not need to care if the other vehicles are autonomous or not. Achieving this goal would improve the efficiency in traffic scenarios with both artificial and human players, as studied in~\cite{gupta2018negotiation} in case of vehicle-pedestrian negotiation, as well as increase people's trust in autonomous vehicles. We believe that seeing self-driving cars hesitating and interpreting the situation, as human usually do to negotiate, would help in breaking the diffidence of the community and support a seamless integration in regular traffic.

Typical solutions~(\cite{urmson2008boss}, \cite{cosgun2017gomentum}) for handling those particular maneuvers consist on rule-based methods which use some notion of the \textit{time-to-collision}~(\cite{vanderhorst1994ttc}), so that they will be executed only if there is enough time in the worst case scenario. These solutions lead to excessively cautious behaviors due to the lack of interpretation of the situation, and suggested the use of machine learning approaches, such as Partially Observable Markov Decision Processes~(\cite{liu2015situation}) or Deep Learning techniques (\cite{isele2017navigating}), in order to infer intentions of other drivers. However, training machine learning algorithms of this kind typically requires simulated environments, and so the behavioral simulation of other drivers plays an important role. 

Popular microscopic traffic simulators, such as \textit{Vissim}~\cite{fellendorf2010vissim} and \textit{Sumo}~\cite{krajzewicz2002sumo}, use specific hard-coded rules to control vehicles dynamic based on common traffic laws. 
For example, in uncontrolled intersections where vehicles have to give way to traffic on the right, each vehicle will wait the freeing of the right lane before entering the intersection. At the same time, a vehicle will yield to all cars within a roundabout before entering.

Even if this conduct seems normal, the capability of negotiation between vehicles is missing: an artificial driver will not try to convince other cars of letting it squeeze in, or it will wait forever if another vehicle with the right of way is yielding. In a nutshell, simulated vehicles are not able to break the behavioral rules, a thing that men and women usually do in normal driving scenarios. 

Recent advances in machine learning suggest the use of \textit{Deep Neural Networks}~(\cite{goodfellow2016dlbook}) to achieve complex behaviors, inducing the agent to learn from representations instead of manually writing rules. In particular, the use of \textit{Reinforcement Learning}~(RL,~\cite{suttonbarto2018rlbook}) techniques to train such agents appears to be a proficient path~(\cite{mnih2015humanlevel}). RL deals with how an agent should act in a given environment in order to maximize the expected cumulative sum of rewards it will receives, with the idea of behaving optimally in the long run instead of chasing the  highest immediate reward. This framework is appealing because it does not need explicit supervision at every decision instant, but the agent learns through a scalar reward signal that reflects the overall result of the actions taken. This reward can also be given once it is known if the whole series of actions led to a good or bad result, like when the car succeed in completing the maneuver or it collides with another vehicle.

A well-known downside of Deep Learning is that it is data-hungry~(\cite{norvig2009data}), namely many examples are needed to robustly train the model. Moreover RL requires both positive and negative experiences, making it unsuitable to be used in real world where negative experiences are generally very expensive to get. These two limitations make the use of a synthetic environment fundamental when using \textit{Deep Reinforcement Learning} (\textit{DRL}).

The aim of this work is to develop a microscopic traffic simulator in which vehicles are endowed with interaction capabilities. This simulator has been thought to be used to train modules able to perform high-level maneuvers with implicit negotiation skills; however it may turn useful for traffic safety and flow analyses as well. Longitudinal accelerations of each vehicle are modeled by a neural network.

Our contributions to reach the goal is twofold: firstly we extend the \textit{Asynchronous Advantage Actor-Critic} (\textit{A3C},~\cite{mnih2016async}) to operate in a multi-agent setting; secondly, we propose a novel architecture which permits to modify the behavior of every agent, even if they all share parameters of the same network as required by the A3C scenario. Agents are collectively trained in a multi-agent fashion so that cooperative behaviors can emerge, gradually inducing agents to learn to interact with each other.

Each agent receives as input a sequence of images representing the scene portions that can be seen at different times translated as top views, taking a cue from the ATARI environment. Using a visual input makes the system ready to be applicable to different scenarios without the need of having a situation-dependent input. However, this kind of input alone does not give the possibility of differentiate the behavior between different drivers.

Because of that, the introduced architecture permits to incorporate the visual input with a vector of parameters, which are used to tune the behavior of the driver. Indeed, every human has its own peculiar way of driving affected by its temper and external factors resulting in impatience or calmness depending on the situation: let's think for this purpose on how we drive when we are in a hurry or when we are carrying a baby to the kindergarten. Encapsulating drivers with different styles inside the simulator makes the simulation more realistic and would lead to the development of more robust applications.
Moreover, the parameter input can be used to give additional details which are not perceivable from images, such as the absolute speed of the agent, enriching the incoming information to the network.

\begin{figure}[htbp]
	\centering
	\subfloat[Real world]{
		\includegraphics[width=0.45\columnwidth]{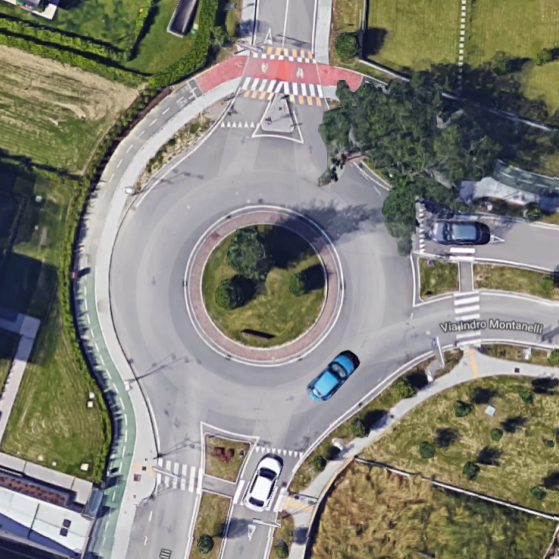}
		\label{fig:real_round}
	}
	\hspace{0.0em}	
	\subfloat[Synthetic representation]{
		\includegraphics[width=0.45\columnwidth]{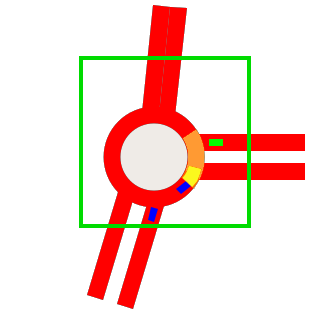}
		\label{fig:synth_round}
	}
	\\
	\vspace{1.0em}
	\subfloat[Navigable]{
		\includegraphics[width=0.2\columnwidth]{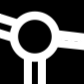}
		\label{fig:nav_space}
	}
	\hspace{1.5em}
	\subfloat[Obstacles]{
		\includegraphics[width=0.2\columnwidth]{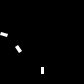}
		\label{fig:obstacles}
	}
	\hspace{1.5em}
	\subfloat[Path]{
		\includegraphics[width=0.2\columnwidth]{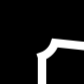}
		\label{fig:path}
	}
	\caption{Synthetic representation of a real world scenario. The top-view of the real scene depicted in~\textbf{\protect\subref{fig:real_round}} is translated in the synthetic representation shown in~\textbf{\protect\subref{fig:synth_round}}. \textbf{\protect\subref{fig:nav_space}}, \textbf{\protect\subref{fig:obstacles}} and  \textbf{\protect\subref{fig:path}} are the semantic layers of the region inside the green square used as visual input for the green agent.}
	\label{fig:representation}
\end{figure}

The proposed solution is evaluated on a synthetic single-lane roundabout scenario shown in Figure~\ref{fig:synth_round}, but the same approach can be adopted in other scenarios as well.
The advantage of abstract representations of this kind is their ease of reproduction with both synthetic and real data, so that high-level maneuver modules trained upon the simulator can be directly used in real world with little effort. A similar approach has been taken concurrently to this work in~\cite{waymo2018chauffeur}.
%\vspace*{6.0em}
\section{Related Works}
Road users simulation is essential to the development of maneuver decision making modules for automated vehicles. In~\cite{isele2017navigating} a system able to enter in an intersection is trained while other vehicles followed a deterministic model called \textit{Intelligent Driver Model} (IDM,~\cite{treiber2000idm}).
In~\cite{liu2018elements} a lane change maneuver module is learned using DRL in a scenario where other vehicles follow a simple lane keeping behavior with collision avoidance, while in~\cite{li2015reinforcement} they are also able to overtake relying on hard-coded rules.
In~\cite{paxton2017combining} both high level (maneuver decision) and low level policies (acceleration and braking) are learned for addressing an intersection in which other vehicles behave aggressively following some preprogrammed rules. 

In all these cases, vehicles populating the environment have a similar conduct and their driving styles are undifferentiated, weakening the realism of the environment.

In~\cite{hoel2018automated} the driving styles diversity is increased by assigning randomly generated speed trajectories to the IDM model; in~\cite{liu2015situation} the behavior variation is enforced by adding a finite number of motion intentions which can be adopted by the drivers. 

However, none of the aforementioned works explicitly consider agents with complex behaviors such as negotiation, since their motion depends on some specific features, e.g. the distance from the vehicle ahead. 

Non-trivial vehicles' motion is obtained in~\cite{shwartz2016safe} by imitation learning~(\cite{bojarski2016endtoend}) using data collected from human drivers; this solution is clearly expensive and directed to a specific situation. 

Having a visual input enhance the system capability of modeling complex behaviors and makes it adaptable to a varieties of different problems and road scenarios without the need of shaping the input for each situation. This was initially proved in~\cite{mnih2015humanlevel}, where the same algorithm, network architecture and hyperparameters were used to learn how to play many different \textit{ATARI 2600} games. The difficulty posed by the correlation between subsequent states in RL was overcome by storing past experiences and updating the agent with mini-batches of old and new examples, a mechanism called \textit{experience replay}.

The system has been improved in~\cite{mnih2016async} increasing its time and memory efficiency thanks to the parallel execution of multiple agents in place of experience replay. In particular, we extended the \textit{Asynchronous Advantage Actor Critic} (\textit{A3C}) so that it can operate in a multi-agent scenarios where many agents are allowed to interact.

In~\cite{gupta2017cooperative} the algorithm called Trust Region Policy Optimization (TRPO~\cite{ schulman2015trpo}) and A3C are extended to multi-agent systems where all agents share parameters of a single policy and contribute to its update simultaneously. In the same work this approach has been evaluated for other two popular DRL algorithms, DQN and Deep Deterministic Policy Gradient (\textit{DDPG}~\cite{lillicrap2015ddpg}), which both rely on experience replay; the results were not satisfying, probably due to the incompatibility between experience replay and the non-stationarity of the system coming from the constantly changing policies of agents populating the environment. 
This suggests that using parallel actor-learners as in A3C has a stabilizing effect which holds in case of a multi-agent setting, since it removes the dependency on the experience replay mechanism. 
A different approach in case of independent learners with no parameter sharing is taken in~\cite{lowe2017multiagent}, where the environment is made stationary by feeding the critic of the DDPG algorithm at training time with actions and observations of all agents. 

Parameter sharing is appealing but forces the agents to behave similarly. For this reason we coupled the visual input with some parameters whose values influence the agent policy, inducing different and tunable behaviors. 
The idea of mixing a high-dimensional visual input and a low-dimensional stream was taken from~\cite{vladlen2017learning}, in which the low-dimensional vector, called \textit{measurements}, is used to define the goal for the agent and to drive the setting toward supervised learning by learning how actions modify future values of the \textit{measurements}. However, this solution fits those problems that have a natural concept of relevant and observable set of measurements (such as video-game playing), but fails on problems in which only a sparse and delayed feedback is available. Hence, we used the low-dimensional input just as an additional input, and followed standard temporal-difference reinforcement learning for training our agents.
%\vspace*{6.0em}
\section{Background}

\subsection{Reinforcement Learning}
In this work we used Reinforcement Learning methods applied to the \textit{Markov Decision Process} framework, in which an agent takes actions at discrete time steps based on the state of the environment. At every step the agent receives a reward signal together with the new state of the updated environment, which is modified by both its natural progress and the action of the agent itself.
The reward signal is generally a simple scalar value and has a central role in the learning process, since it is the only way for the agent to evaluate its actions. This introduces a well-known difficulty of RL: the feedback is generally the result of many actions, possibly taken several steps in the past.

Describing the process in mathematical terms, the action at time \textit{t} is represented by $a_t$ and is taken from a discrete or continuous range of actions referred to as \textit{A}; the state at time \textit{t} is represented as $s_t$ while the reward as $r_t$.
The goal of the agent is to learn a policy $\pi(a|s)$, namely the probability of taking action \textit{a} given the state \textit{s}, in order to maximize the expectation of a function of future rewards called \textit{return}. The return is generally defined as $R_t = \sum_t^T r_t + \gamma r_{t+1} + ... + \gamma^{T-t} r_{T-t}$, where T is the terminal time step and $\gamma$ is a discount factor used to modulate the importance of future rewards.

Useful for the estimation of a policy are the \textit{value functions}. The \textit{state-value function}~(\ref{eq:state_value_function}) of a policy $\pi$ gives the expected return when, starting from state \textit{s}, the policy $\pi$ is followed; the \textit{action-value function}~(\ref{eq:action_value_function}), similarly, gives the expected return if we follow policy $\pi$ after taking action \textit{a} in state \textit{s}.
\begin{equation}\label{eq:state_value_function}
V_\pi(s_t) = \mathbb{E}(R_t|s_t) 
\end{equation}
\begin{equation}\label{eq:action_value_function}
Q_\pi(s_t,a_t) = \mathbb{E}(R_t|s_t,a_t)
\end{equation}
If the agent was able to find the \textit{optimal action-value function} $Q^*(s,a)$, namely the action-value function of the best policy, the problem would be solved since it could pick the best action from a generic state $s_t$ simply by choosing $a^* = \max_{a \in A} Q^*(s,a)$. 

Trying to estimate this function is the goal of those algorithms which fall under the category of \textit{action-value methods} such as \textit{Q-learning} (\cite{watkins1989qlearning}). If we approximate $Q^*(s,a)$ with a function $q(s,a;\theta)$ modeled by the vector of parameters $\theta$, in \textit{1-step Q-learning} the parameters are updated as:
\begin{equation}\label{eq:action_value_update}
\theta_{t+1} = \theta_t + \alpha  \cdot K \frac{\partial q(s_{t},a_{t};\theta_t)}{\partial \theta_t}
\end{equation}
where $\alpha$ is a parameter dependent on the optimization algorithm while \textit{K} tells if the approximated action-value function should be increased or decreased around the pair $\{s_t, a_t\}$ and is defined as:
\begin{equation}\label{eq:action_value_advantage}
K = r_t + max_{a^*}q(s_{t+1},a^*;\theta_t) - q(s_{t},a_{t};\theta_t)
\end{equation}
It is worth noting that the target value $r_t + max_{a^*}q(s_{t+1},a^*;\theta_t)$ is an estimate itself because it uses the estimated action-value function of the next state as better approximation of the real optimal action-value. This operation of learning \lq\lq a guess from a guess\rq\rq\ is called \textit{bootstrapping}.

A different approach is taken in \textit{policy-based methods}, which directly estimate the policy instead of a value function. In this category falls the algorithm called \textit{Reinforce} (\cite{williams1992reinforce}) in which parameters are updated such that probabilities of actions that yielded a better cumulative reward at the end of an episode are increased with respect to the lower-return actions. Approximating the optimal policy $\pi^*(a|s)$ with a function $\pi(a|s; \theta)$, the updates are:
\begin{equation}\label{eq:reinforce_update}
\theta_{t+1} = \theta_t + \alpha \cdot R_t \frac{\partial \log \pi(a|s; \theta)}{\partial \theta}
\end{equation}
In this algorithm the updates can be computed only at the end of an episode, since the true cumulative reward is needed.

Between action-value and policy-based methods there is an hybrid family of algorithms called \textit{actor-critic methods}.

In this setting the policy is directly optimized as in policy-based solutions; however, the value function is also estimated, giving two benefits: firstly the value of a state is used to reduce the variance of updates, as explained in~\cite{suttonbarto2018rlbook}; furthermore the estimated state-value function permits bootstrapping, thus avoiding the need to wait for the end of an episode before performing a parameters update. The probability of actions can now be modulated with the following \textit{Advantage} $A_b$ instead of the full reward $R_t$:
\begin{equation}\label{eq:actor_critic_advantage}
A_b = r_t + \gamma r_{t+1} + ... + \gamma^{b-1} r_{t+b-1} + v(s_{t+b} ; \theta^v) - v(s_t ;\theta^v)
\end{equation}
where $b$ is the number of real rewards used before bootstrapping, called \textit{bootstrapping interval}.

Calling the approximated optimal policy $\pi(a|s; \theta^\pi)$ and the state-value function $v(s; \theta^v)$, the updates of this two functions can now be defined as:
\begin{equation}\label{eq:actor_update}
\theta^\pi_{t+1} = \theta^\pi_{t} + \alpha_\pi \cdot A_b \frac{\partial \log \pi(a|s; \theta^\pi)}{\partial \theta^\pi}
\end{equation}
\begin{equation}\label{eq:critic_update}
\theta^v_{t+1} = \theta^v_t + \alpha_v \cdot A_b \frac{\partial v(s_t; \theta^v)}{\partial \theta^v}
\end{equation}
If the bootstrapping interval $b$ equals 1, only the reward following an action is used to directly evaluate that action. This setting is ideal in scenarios where the complete consequences of an action are known immediately (e.g. face correctly recognized by a face detection agent); however, it is not optimal in situation where the reward is delayed (e.g. agent regulating the water level of a dam). Increasing $b$, a longer series of rewards will be accumulated before estimating the value function; in this way delayed rewards are propagated faster but their merit is divided among all the actions performed in the time window.

In the \textit{n-step actor critic} algorithm used in A3C a mixed approach is adopted. In this solution both long and short-term bootstrapping take place: every $n$ time-steps (or if a terminal state is reached) a sequence of updates like those in equations~(\ref{eq:actor_update}) and~(\ref{eq:critic_update}) are  performed, one for each time-step. In each update the longest possible series of rewards is used for estimating the value of the state, ranging from a 1-step to n-step updates. This process is shown in Algorithm~\ref{alg:nstep}.

\begin{algorithm}[]
	\caption{\textit{n-step} Advantage Actor Critic \label{alg:nstep}}
	\begin{algorithmic}[1]
		\STATE initialize parameters $\theta_\pi$ of $\pi(a|s; \theta_\pi)$
		\STATE initialize parameters $\theta_v$ of $V(a|s; \theta_v)$
		\STATE $t \leftarrow 1$
		\WHILE{ $learning \; is \; active$ }
		\STATE get $s_t$
		\STATE $t_{last\_up} = t$
		\WHILE{ $t - t_{last\_up} < n$ \textbf{and} $s_t \; is \; not \; terminal$ }
		\STATE execute action $a_t$
		\STATE get reward $r_t$, state $s_{t+1}$
		\STATE $t \leftarrow t+1$\\
		\ENDWHILE
		\STATE $R=$
		$\begin{cases}
		0 & \text{if $s_t$ is terminal} \\
		V(s_t; \theta_v) & \text{otherwise}		
		\end{cases}$
		\STATE $\Delta_\pi = 0, \; \Delta_v = 0$
		\FORALL{ $i = t-1, .. , t-t_{last\_up}$ }
		\STATE $R = r_i + \gamma r$
		\STATE $\Delta_\pi = \Delta_\pi + \nabla_{\theta_\pi} \log \pi(a_i|s_i; \theta_\pi) (R - V(s_i; \theta_v))$
		\STATE $\Delta_v = \Delta_v + \nabla_{\theta_v} (R - V(s_i; \theta_v))^2$
		\ENDFOR
		\STATE get $\alpha_\pi, \; \alpha_v$ from the optimization algorithm
		\STATE $\theta_\pi = \theta_\pi + \alpha_\pi \cdot \Delta_\pi$
		\STATE $\theta_v = \theta_v + \alpha_v \cdot \Delta_v$
		\ENDWHILE
	\end{algorithmic}
\end{algorithm}

\subsection{Asynchronous Advantage Actor Critic - A3C}
A problem of using Deep Neural Networks as function approximators in RL is the correlation between updates that comes from the sequentiality of data in the process. 

This correlation was initially broken by using the so called experience replay~(\cite{mnih2015humanlevel}): \textit{tuples}~$\{s_t, a_t, s_{t+1}, r_t\}$ were stored instead of being used immediately, while the parameters updates were computed using randomly sampled tuples from the replay buffer. In this setting, updates are not based on sequential data anymore, but the replay buffer makes the process expensive in terms of memory occupation and it forces the use of old data.

An improved solution, adopted in~\cite{mnih2016async}, consists in running several actor-critic agents in parallel, each one with its own copy of the parameters of a global neural network.
Every agent acts in a different instance of the environment and sends its updates, computed as in Algorithm~\ref{alg:nstep}, asynchronously from the other agents. Since each agent is seeing a different environment configuration, their updates are not correlated and contribute in augmenting the stability of the process.
A visual representation of this system is shown in Figure~\ref{fig:single_agent}.
%\vspace*{3.0em}
\section{Microscopic Traffic Simulation}\label{framework}%multi agent domain - microscopic simulation ....

\subsection{Multi-agent A3C}
In the normal A3C setting, multiple agents act in parallel in independent instances of the same environment. However, it is not possible for them to learn the interaction with each other, since each agent is sensing a state which is independent from that of the others. 

In order to make this interaction possible we let several agents share the same playground, so that they will be able to sense each other. For instance, let's consider agents $a$ and $b$ acting on the same environment: $a$, by taking action $a_t^a$ at time step $t$ based on policy $\pi(a_t^a | s_t^a)$, could affect state $s_{t+1}^b$ of the agent $b$. If this happens, $b$ will react accordingly taking action $\pi(a_{t+1}^b | s_{t+1}^b)$, potentially affecting in turn $s_{t+2}^a$. Therefore, learning the optimal policy $\pi$ for an agent means taking into account not only how the environment is affected by its actions, but also what behavior they induce to other agents.

In practice, this implies that agents have to wait that the others took their actions before receiving the subsequent state. This affects negatively the performance of the algorithm, since, due to the asynchronous nature of the process, some of the agents could be computing their updates, that is executing a time consuming backpropagation. Moreover, this solution jeopardizes the stability of the system because parameter updates of agents acting in the same environment are not independent anymore. For this reason, in multi-agent A3C we run several environment instances as in traditional A3C, each one populated with several active agents.

\begin{figure}[t]
	\centering
	\subfloat[Single-agent setting]{
		\includegraphics[width=0.9\columnwidth]{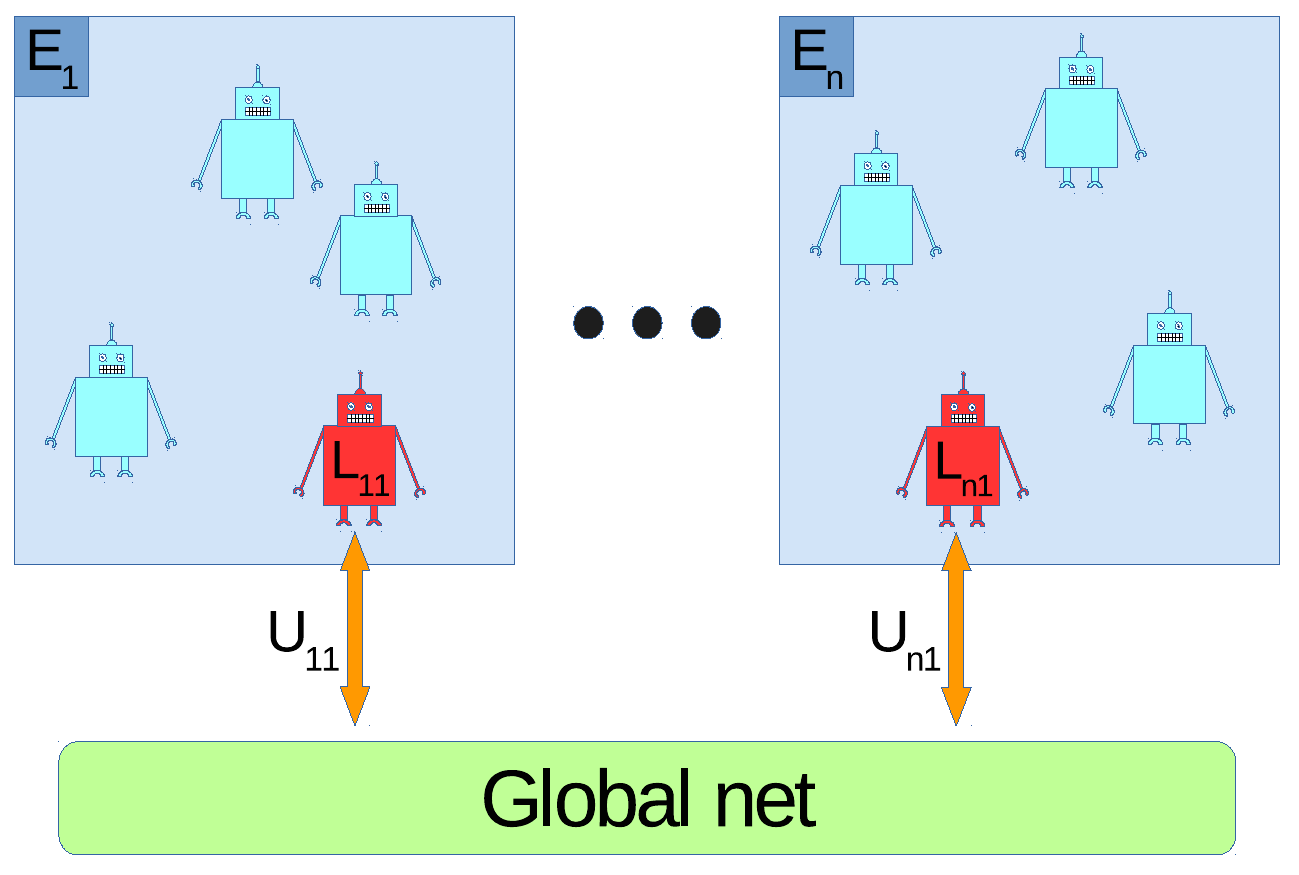}
		\label{fig:single_agent}
	}
	\\
	\vspace{0.5em}
%	\hspace{0.2em}
	\subfloat[Multi-agent setting]{
		\includegraphics[width=0.9\columnwidth]{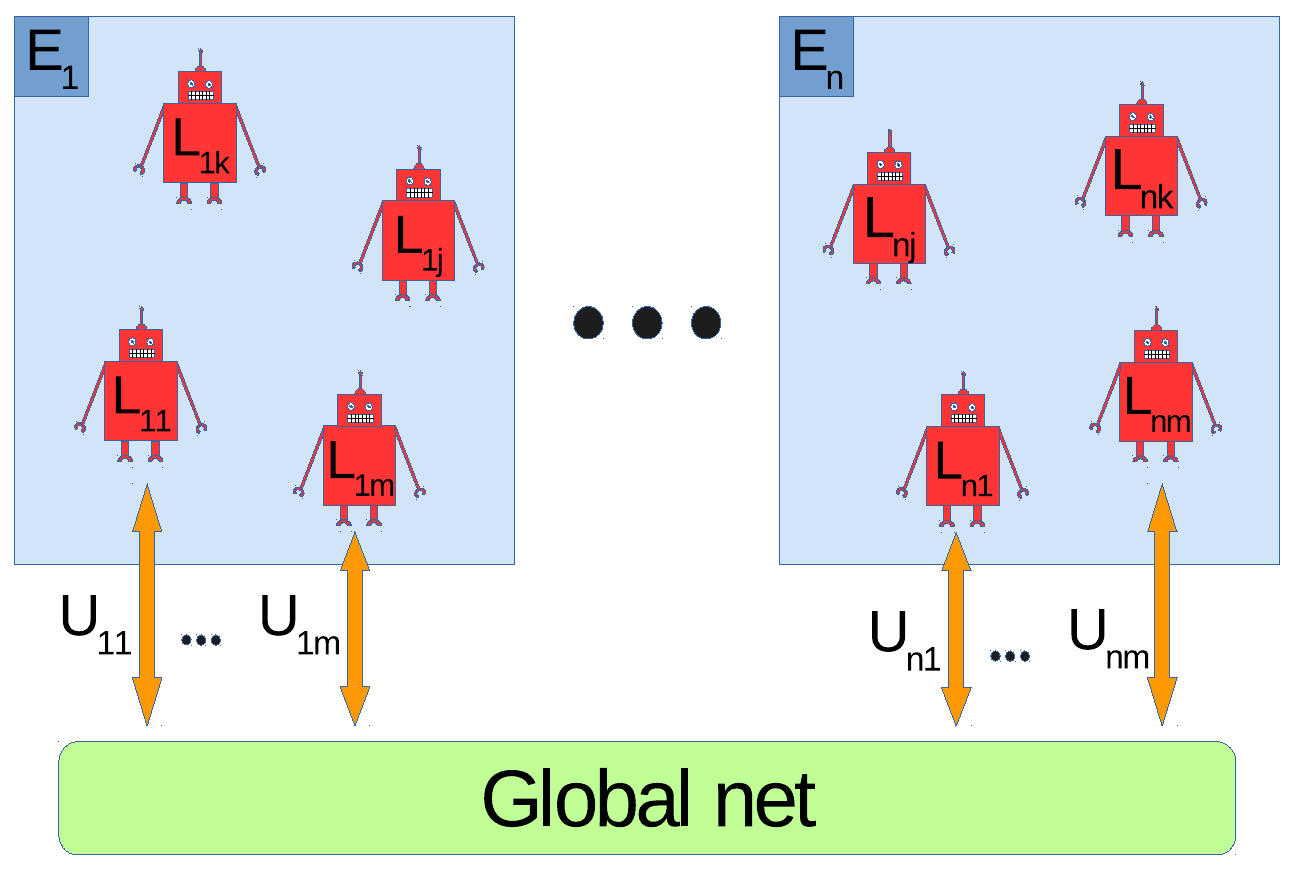}
		\label{fig:multi_agent}
	}
	\caption{Comparison between single and multi-agent A3C settings. Red robots are learning agents acting in environment $E_i$, while blue robots are passive agents. Each active agent owns a copy $L_j$ of the global network, and it contributes to its evolution by updates $U_j$. In~\textbf{\protect\subref{fig:single_agent}} learning agents play in an environment populated only by non-active agents, while in~\textbf{\protect\subref{fig:multi_agent}} active agents can sense each other, allowing them to learn how to interact.}
	\label{fig:results1}
\end{figure}

Each agent accumulates updates every \textit{n} frames, and sends them to a global copy of the network only at the end of the episode in order to reduce the synchronization burden, updating at the same time its local copy with the updates coming from the other agents. Since some of those agents are accumulating experience in different environment instances, the correlation between updates is diminished, making the process more stable. Algorithm~\ref{alg:multi_nstep} shows the process of multi-agent A3C, whose visual representation is given in Figure~\ref{fig:multi_agent}.

In our experiment we set $n = 20$ and $\gamma = 0.99$; we used \textit{RMSProp} optimizer with a decay factor of $\alpha = 0.99$ and initial learning rate of $7e^{-4}$.

\begin{algorithm}[]
	\caption{multi-agent \textit{n-step} Advantage Actor Critic \label{alg:multi_nstep}}
	\begin{algorithmic}[1]
		\STATE initialize parameters $\theta_\pi^g$ of $\pi(a|s; \theta_\pi)$
		\STATE initialize parameters $\theta_v^g$ of $V(a|s; \theta_v)$
		\STATE get $n_{env}$ as the number of environment instances
		\STATE get $n_{ag}$ as the number of agents for each environment
		
		\STATE \textbf{run} $n_{env}$ environment instances concurrently
		\FORALL{$environment \; instance$}
		\STATE \textbf{run} $n_{ag}$ threads concurrently
		
		\ENDFOR
		
		\FORALL[\textit{executed concurrently}]{$agent \; a$}
		\STATE $t \leftarrow 1$
		\WHILE{$training \; is \; active$}	
		\STATE copy parameters $\theta_\pi^a = \theta_\pi^g$
		\STATE copy parameters $\theta_v^a = \theta_v^g$
		\STATE $\Delta_\pi^a = 0, \; \Delta_v^a = 0$
		\WHILE{$s_t \; is \; not \; terminal$}
		
		\STATE get $s_t$
		\STATE $t_{last\_up} = t$
		\WHILE{ $t-t_{last\_up}<n$ \textbf{and} $s_t \; is \; not \; terminal$}
		\STATE execute action $a_t$ from $\pi(a_t|s_t;\theta_\pi^a)$
		\STATE wait other agents to finish their actions
		\STATE get reward $r_t$, state $s_{t+1}$
		\STATE $t \leftarrow t+1$
		\ENDWHILE
		\STATE $R=$
		$\begin{cases}
		0 & \text{if $s_t$ is terminal} \\
		V(s_t, \theta_v^a) & \text{otherwise}		
		\end{cases}$
		\FORALL{ $i = t-1, .. , t-t_{last\_up}$ }
		\STATE $R = r_i + \gamma r$
		\STATE $\Delta_\pi^a = \Delta_\pi^a + \nabla_{\theta_\pi^a} \log \pi(a_i|s_i; \theta_\pi^a) (R - V(a|s_i; \theta_v^a))$
		\STATE $\Delta_v^a = \Delta_v^a + \nabla_{\theta_v^a} (R - V(a|s_i; \theta_v^a))^2$
		\ENDFOR
		\ENDWHILE
		\STATE \textbf{update} $\theta_\pi^g$ using $\Delta_\pi^a$ %\tikzmark{top} \tikzmark{right}
		\STATE \textbf{update} $\theta_v^g$ using $\Delta_v^a$% \tikzmark{bottom}
		\ENDWHILE 
		\ENDFOR
	\end{algorithmic}
	%\AddNote{top}{bottom}{right}{\textit{Executed atomically}}
\end{algorithm}

Depending on how the reward function is shaped, the learning process may be cooperative or competitive~(\cite{tampuu2015multiagent}). The former is the way we followed to simulate behavior of road vehicles:  a positive reward is given to the artificial driver if it reaches its goal while a negative reward is given to those drivers involved in a crash. Agents are stimulated to reach their goals avoiding other agents, leading to an implicit negotiation when the paths of different vehicles intersect.

We think that this multi-agent learning setting is captivating for many applications that require a simulation environment with intelligent agents, because it learns the joint interactive behavior eliminating the need for hand-designed behavioral rules. %The environment acquires consistency throughout the training phase while every agent is learning to play.

\subsection{Input architecture}\label{input_architecture}
Having a visual input organized as images is appealing since it frees from the need of defining a case-specific design for each different scenario. Moreover, it frees from the need of considering a fixed amount of vehicles~(as~in~\cite{hoel2018automated}).

However, this kind of input has a drawback: the convolutional pipeline makes it not suitable for communicating numerical information, such as the desired behavior for the agent or its precise speed. Transferring this type of information is possible using fully-connected layers, but it becomes inefficient to directly process visual images by this mean. For this reason, we shaped our system in order to have a mixed input, made of \textit{visual} and \textit{numerical} information, which is processed by two different pipelines as elucidated in Section~\ref{network_architecture}. 

\subsubsection{Visual input}
The agent may not be able to sense the whole environment but just a portion of it, which we call its \textit{view}.
The visual part of the input consists on a historical sequence of the $v$ most recent views seen from the agent. We set $v = 4$ as in~\cite{mnih2015humanlevel} making the estimation of relative speed and acceleration feasible. Each view is divided in semantic layers separating information with different origin. As explained in Section~\ref{state_space} we included semantic layers related to the navigable space, the path the agent will follow and obstacles around the vehicle; however, several other layers could be added, such as the occluded area or the road markings, in order to enrich the input information.

\subsubsection{Numerical input}
The numerical component permits the augmentation of the input with information which are not directly perceivable from the sequence of views, such as the absolute speed of the agent or the exact distance which is still to be traveled to reach the goal. This allows a better understanding of the scene, i.e. a better estimation of the states value.
More importantly, the numerical input can be used to shape agents attitude by teaching it a direct correlation between some of its tunable input and the to-be-obtained reward. Practically, they can be used to partially perturb the state of the agent at will. 
In this work we took advantage of numerical input to the network for tuning the agent aggressiveness and to suggest a desired cruising speed, as explained in Sections~\ref{aggressiveness} and~\ref{targetspeedtuning} respectively.

\subsection{Network architecture}\label{network_architecture}
Because of the different nature of the two input streams, they are initially processed by two different pipelines, aimed to produce two feature vectors. 

The image processing pipeline is constituted by two convolutional and one fully connected layers, whereas the numerical input vector is processed by two fully connected layers. Each pipeline produces a feature vector as output. The two output vectors are joined together before being handled by the last fully-connected hidden layer. After every transformation a \textit{ReLU} nonlinearity is applied to the output.

The idea of this hybrid architecture has been inspired from~\cite{vladlen2017learning}, in which an input made of a single image and two numerical vectors is processed by a network built in a similar fashion.

\begin{figure}[]\centering
	\includegraphics[width=1.0\columnwidth]{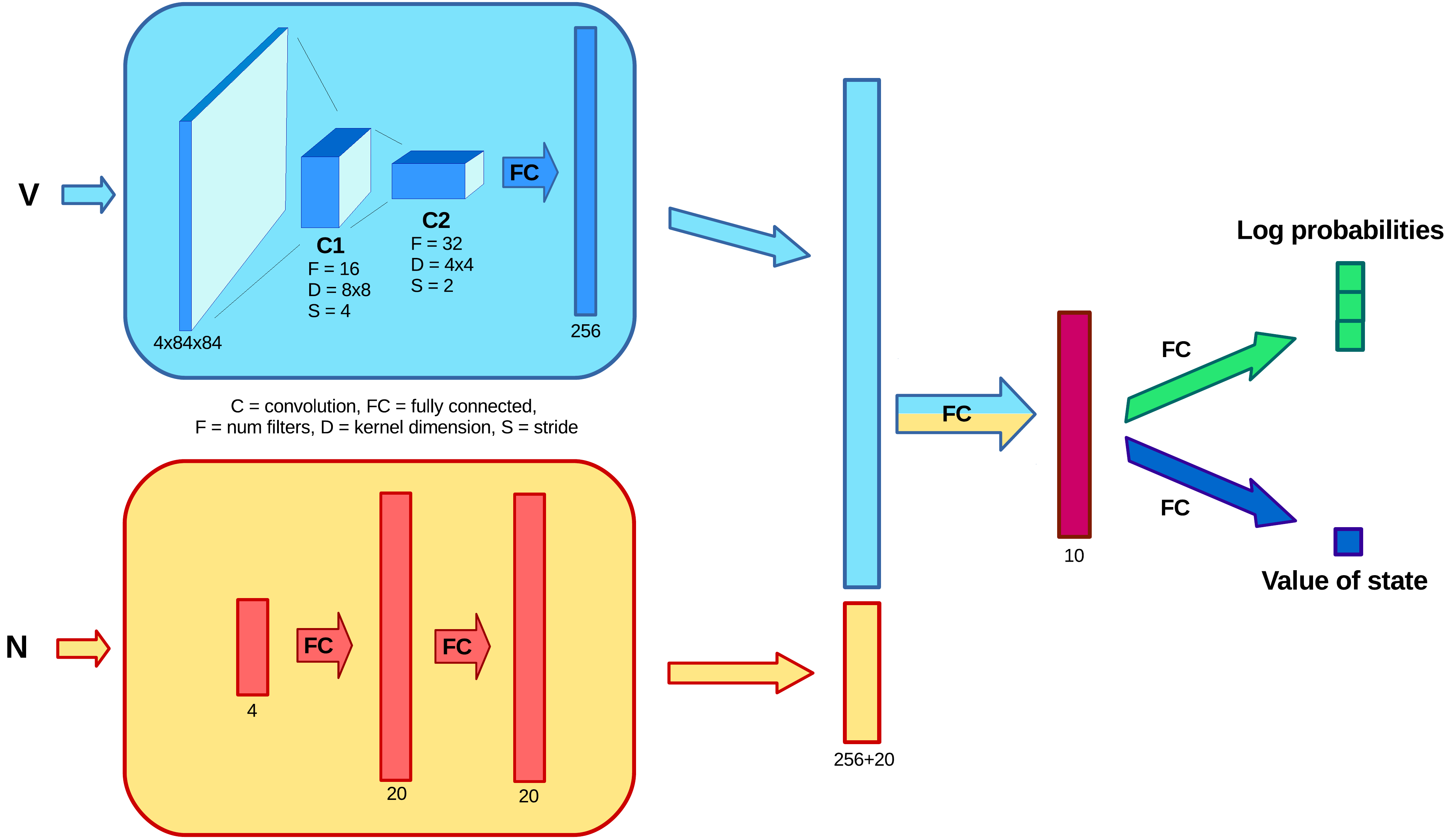}
	\caption{Architecture of the network. Visual input~\textbf{V} and numerical input~\textbf{N} initially follow two different pipelines, before being merged in a single vector, which is then used to compute probabilities of actions and to estimate the value of the state.\label{fig:network_architecture}}
\end{figure}

The output of the last hidden layer is finally used for the actions log-probabilities computation and the state-value estimation.
The network architecture is represented in Figure~\ref{fig:network_architecture}. 

%\vspace*{1.0em}
\section{Experiment}
The model explained in Section~\ref{framework} has been used to simulate the traffic of cars on a roundabout with three entries~(Figure~\ref{fig:synth_round}), which is a representation of an existing roundabout~(Figure~\ref{fig:real_round}) made with the help of the \textit{Cairo} graphics library (\cite{cairographics}).

Each vehicle populating the roundabout has its own path which simply depends on a randomly assigned exit; the goal of the agent is to reach the end of its exit lane.
Furthermore, to each agent is assigned a desired cruising speed consisting to the maximum speed the agent is aiming for, which it should be reached and maintained when caution is not needed; this speed will be called from now on its target speed. A new vehicle with a random target speed is spawned from an entry lane dependently on the positions of the other cars, inducing different traffic conditions. 

During the experiments the roundabout was populated by a maximum of six vehicles such that no traffic jams were created, thus making the effects of the agent's behavior tuning appreciable. However, this framework can be used to simulate vehicles' behavior even in high traffic conditions.

\subsection{State space}\label{state_space}
As explained in Section~\ref{input_architecture} the input is made of two different streams, a visual and a numerical one. 
The visual part is a representation of the surrounding of the vehicle within a limited range of $50 x 50$ square meters converted to a sequence of images with $84 x 84$ pixels. This input is made of several semantic layers, which in this experiment are:
\begin{enumerate}
	\item \textbf{Navigable space}: the part of the space in which vehicles can drive, generally given from the maps and shown in Figure~\ref{fig:nav_space};
	\item \textbf{Obstacles}: made of all vehicles and obstacles seen from the agent, including the agent himself; this information can be obtained from the perception module of the self-driving vehicle and it is shown in Figure~\ref{fig:obstacles};
	\item \textbf{Path}: that shows to the agent its assigned path, and depends on the choices of the high-level planner; this layer is shown in Figure~\ref{fig:path}.
\end{enumerate}  

The numerical part, on the other hand, consists of several scalar values which contain both additional information about the scene and clues to predict future rewards, which are used at test time to modulate the agent behavior.
In our instance the numerical input is restricted to the following parameters:
\begin{enumerate}
	\item \textbf{Agent speed}. 
	\item \textbf{Target speed}. 
	\item \textbf{Elapsed time ratio}: ratio between the elapsed time from the beginning of the episode and the time limit to reach the goal. 
	\item \textbf{Distance to the goal}: distance to be traveled before reaching the goal.
\end{enumerate}

The network architecture, together with the hyperparameters used for the experiment, are shown in Figure~\ref{fig:network_architecture}.

\subsection{Action space}
The chosen action space is discrete and boils down to three actions: accelerate, brake or maintain the same speed. We set comfortable acceleration values, which are $1 \frac{m}{s^2}$ in case of positive acceleration and $-2 \frac{m}{s^2}$ in case of braking. 
The acceleration command will have its effect only if the speed is under a maximum accepted value.

\subsection{Reward shaping}\label{reward_shaping}
The reward obtained from an agent is made of three main terms:
\begin{equation}
r_t = r_{terminal} + r_{danger} + r_{speed}
\end{equation}

$r_{terminal}$ depends on how the episode ends, and so differs from zero only if $t$ is a terminal state; it equals:
\begin{itemize}
	\item {$+1$}: if the agent reaches its goal;
	\item {$-1$}: if the agent crashes against another agent;
	\item {$-1$}: if the available time expires.
	\item {$0$}: if $t$ is a non-terminal state.
\end{itemize}

$r_{danger}$ is a soft penalization given in situations which are to be avoided, either because the agent made a dangerous maneuver or because it broke some traffic laws.
Calling $d_v$ the distance traveled from the vehicle $v$ in one second, $r_{danger}$ will be:
\begin{itemize}
	\item {$-k_y$}: if the agent fails to yield when entering the roundabout to an already inserted vehicle $v$. This happens when the agent crosses the region in front of $v$ whose length is taken equal to $3 \cdot d_v$, as shown from the orange region in the example of Figure~\ref{fig:synth_round}. 
	\item {$-k_s$}: if the agent $a$ violates the safety distance with the vehicle in front, unless the latter is entering the roundabout and it is cutting in front of the agent. The safety distance in this case is taken equal to $d_a$, and the associated region is depicted in yellow in Figure~\ref{fig:synth_round}.
	\item {$0$}: if none of the above occur.
\end{itemize}

$r_{speed}$ is a positive reward which is maximized when the actual speed of the agent ($s_a$) coincides with the target speed input parameter ($s_t$):
\begin{equation}
	r_{speed} = \begin{cases}
		\frac{s_a}{s_t} \cdot k_p & \textbf{if $s_a \le s_t$}\\
		k_p - \frac{s_a - s_t}{s_t} \cdot k_n & \textbf{if $s_a > s_t$}
	\end{cases}
\end{equation}  

In our experiment we set $k_y = k_s = 0.05$, $k_p = 0.001$, $k_n = 0.03$.

\subsection{Aggressiveness tuning}\label{aggressiveness}
In the training phase, agents learn to increase their speed to reach the goal within a time limit. This is possible because both time left and remaining distance to be traveled are provided as input. 
It was not possible to achieve this same behavior without using the distance-to-the-goal input, highlighting the agent inability to realize from images that increasing its speed the goal will be reached faster. Nevertheless, things work well when this information is enforced with the remaining distance, confirming the power of the visual-numerical coupling.

This experience has been exploited at test time to vary the agent aggressiveness by tuning its time and distance left at will, since it is not needed anymore to limit the episode length. Doing this each agent present in the simulator will have its own unique attitude, resulting in a more heterogeneous scenario.

We explored scenarios in which the elapsed time ratio was replaced by an arbitrary real value in the $[0,1]$ interval, keeping the distance to an imaginary goal fixed.
Values close to 1 will induce the agent to drive faster, in order to avoid the predicted negative return for running out of time. For the same reason, values close to 0 will tell the driver that it still has much time, and it is not a problem to yield to other vehicles. This way, the elapsed time ratio acts as an \lq\lq aggressiveness level\rq\rq\ for the agent.

An experiment has been set up in order to test the validity of this approach. We populated the roundabout with six vehicles having random aggressiveness levels, and we tested how an agent with a fixed known aggressiveness behaves by letting it play several episodes.
We registered two different parameters as feedback of each driver style: the ratio of positive ending episodes, in which the agent reaches the goal, and its average speed. Results are shown in the graph of Figure~\ref{fig:aggressiveness_graph}. It is clearly visible that, by increasing the aggressiveness level, the driver behavior will shift towards a more risky configuration, since both average speed and probability of accidents grow. Moreover, it is interesting to see that values of aggressiveness outside the training interval $[0,1]$ produce consistent consequences to the agent conduct, intensifying its behavior even further.   

%\begin{figure}
%	\includegraphics[width=0.7\columnwidth]{images/grafico_angriness}
%	\caption{Effect of aggressiveness input on the behavior of the agent. The red plot shows the positive-ending episodes ratio varying the aggressiveness input. The blue plot shows how this input affect the average speed along an episode\label{fig:aggressiveness_graph}}
%\end{figure}

\subsection{Target speed tuning}\label{targetspeedtuning}
The $r_{speed}$ reward term, presented in Section~\ref{reward_shaping}, increases until the speed of the agent gets closer to its target speed, and it starts to diminish for agent speeds above the desired one.  
At training time, agents were assigned a random target speed for every episode, taken from a uniform distribution between $[5,8] \frac{m}{s}$, inducing them to learn to not surpass that tunable value and so simulating drivers with different desired speeds. Agents can still decide to surpass their target speed if this will bring advantages, such as stopping the insertion of another vehicle by a particular aggressive driver.

To evaluate how the dedicated input affects the behavior of the agent we let several agents, having different target speed values but equal aggressiveness levels of $0.5$, to play several runs in scenarios similar to the one used to test the aggressiveness tuning in Section~\ref{aggressiveness}. 

\begin{figure}[]
	\centering
	\subfloat[aggressiveness test]{
		\includegraphics[width=0.7\columnwidth]{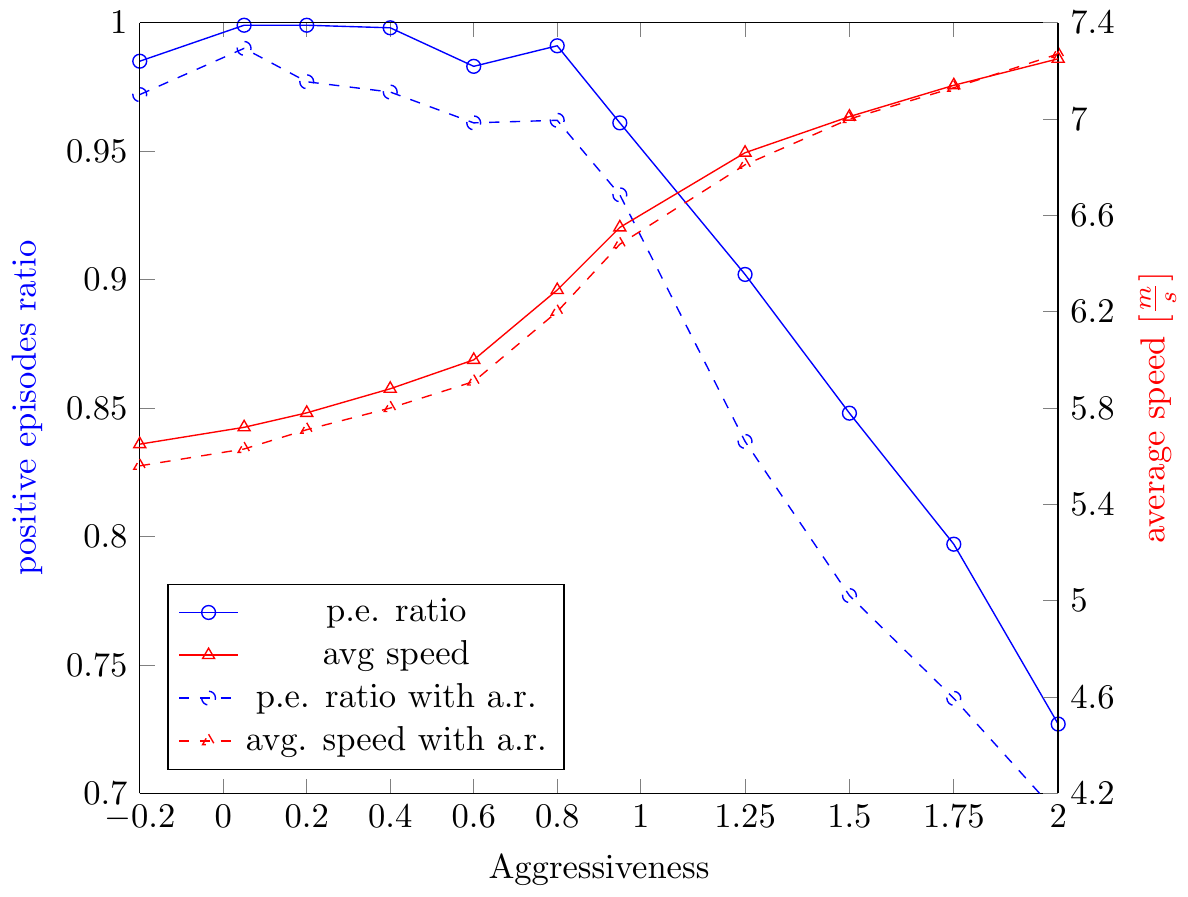}
		\label{fig:aggressiveness_graph}
	}
	//
	\subfloat[speed test]{
		\includegraphics[width=0.7\columnwidth]{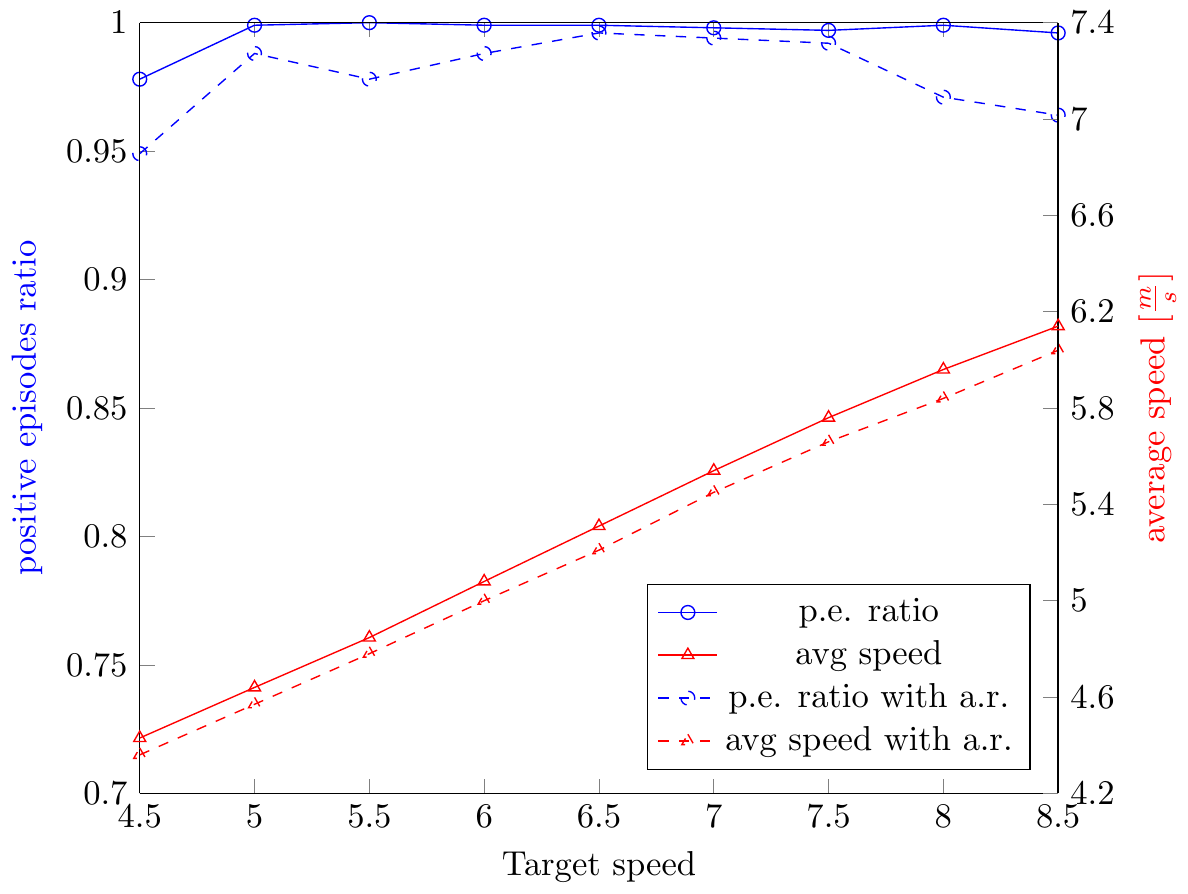}
		\label{fig:speed_graph}
	}
	\caption{Effect of aggressiveness~\protect\subref{fig:aggressiveness_graph} and target speed~\protect\subref{fig:speed_graph} levels variation on the agent behavior. Blue plots show the agent positive-ending episodes ratio while the red plots its average speed. Dashed lines refer to an agent whose action are repeated for 4 times while solid lines to an agent free to choose actions at every time step. Each data point is obtained averaging over 5000 episodes.\label{fig:tests}}
\end{figure}

The graph with the results, given in Figure~\ref{fig:speed_graph}, shows that an increment on the agent target speed leads to an higher average speed, without affecting negatively the positive ending ratio. Even in this case, values of the target speed outside the training range $[5,8]$ induce consistent agent behaviors.

%\begin{figure}
%	\includegraphics[width=0.7\columnwidth]{images/grafico_speed}
%	\caption{Effect of target speed input on the behavior of the agent. The red plot shows the positive-ending episodes ratio varying the target speed. The blue plot shows how this input affect the average speed along an episode\label{fig:speed_graph}}
%\end{figure}

\subsection{Environment settings comparisons}
We compared agents trained with 1 and 4 environment instances in a simplified case in which the target speed given to the agent cannot be exceeded (once reached the target speed the accelerate command has no effect), in order to speed up learning for comparison purposes. We tested both with and without \textit{action repeat} of 4 (\cite{mnih2015humanlevel}), that is repeating the last action for the following 4 frames (repeated actions are not used for computing the updates). It has been proved that in some cases action repeat improves learning (\cite{braylan2015frameskip}) by increasing the capability to learn associations between temporally distant state-action pairs, giving to actions more time to affect the state. 

\begin{figure}[]
	\centering
	\subfloat[1 instance, yes action repeat]{
		\includegraphics[width=0.5\columnwidth]{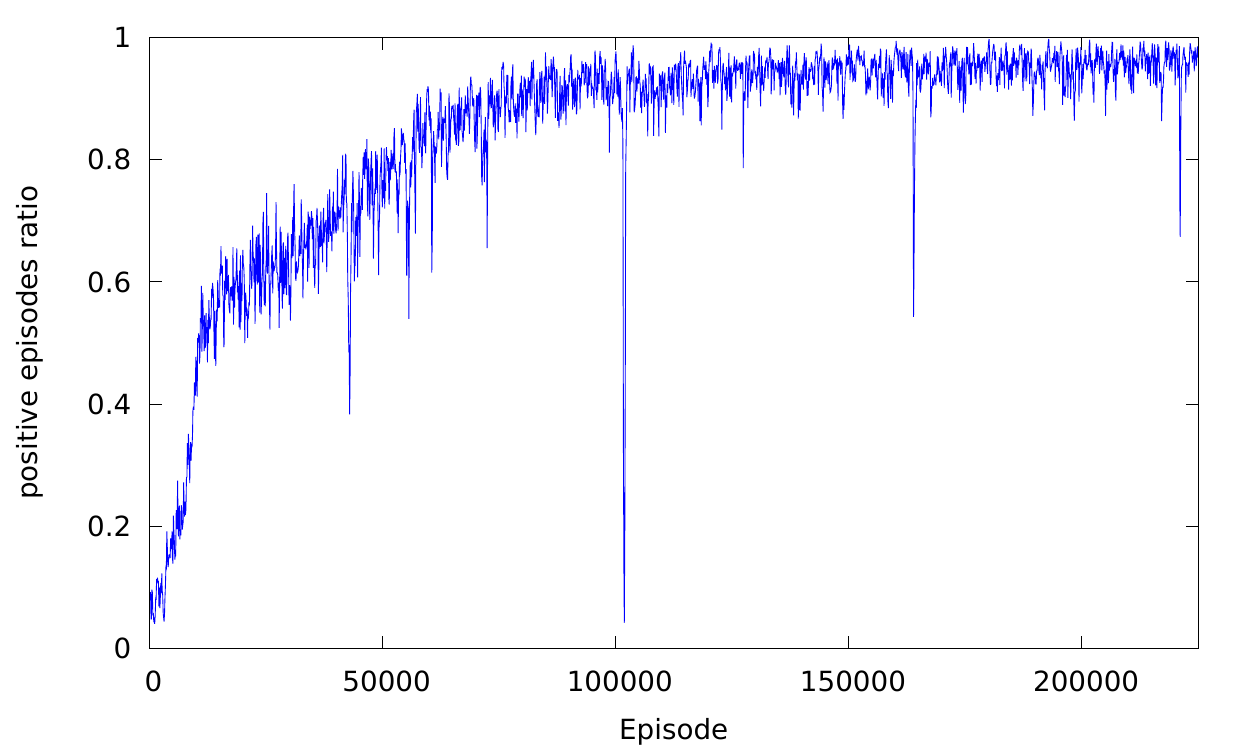}
		\label{fig:1_ar}
	}
%	\hspace{0.1em}
	\subfloat[4 instances, yes action repeat]{
		\includegraphics[width=0.5\columnwidth]{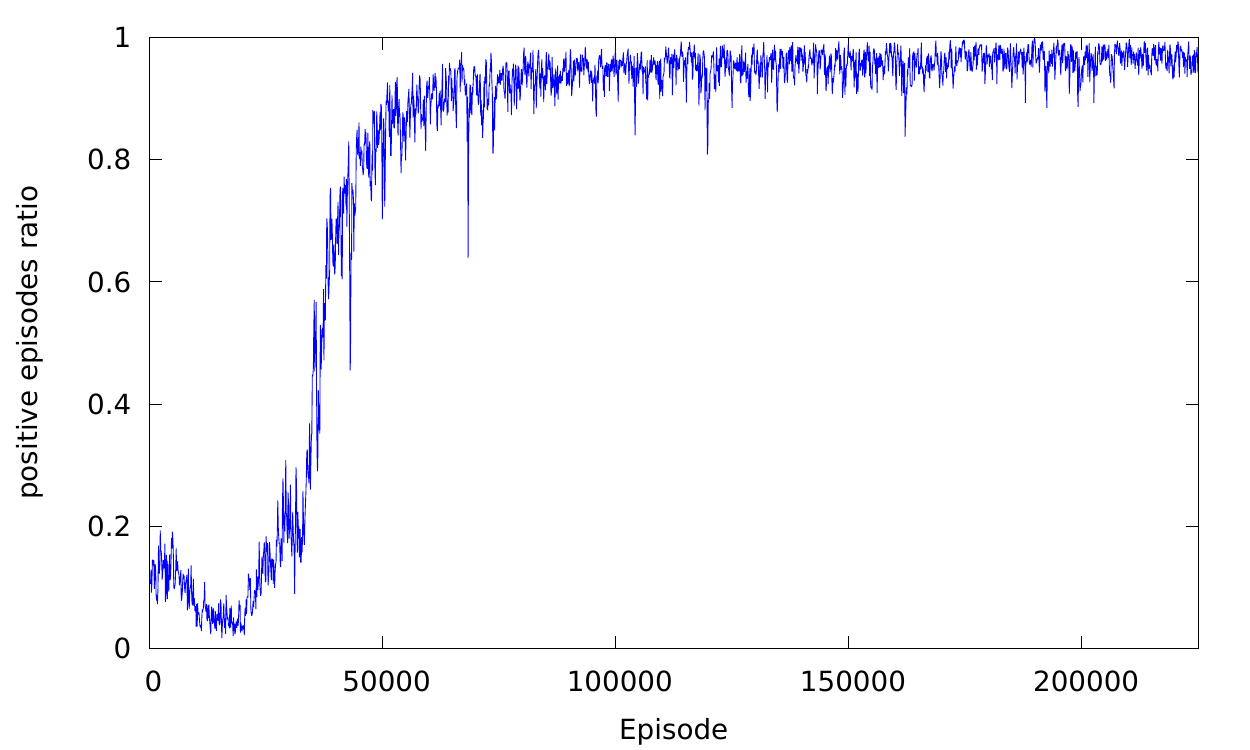}
		\label{fig:4_ar}
	}    
	\\
	\vspace{0.5em}
	\subfloat[1 instance, no action repeat]{
		\includegraphics[width=0.5\columnwidth]{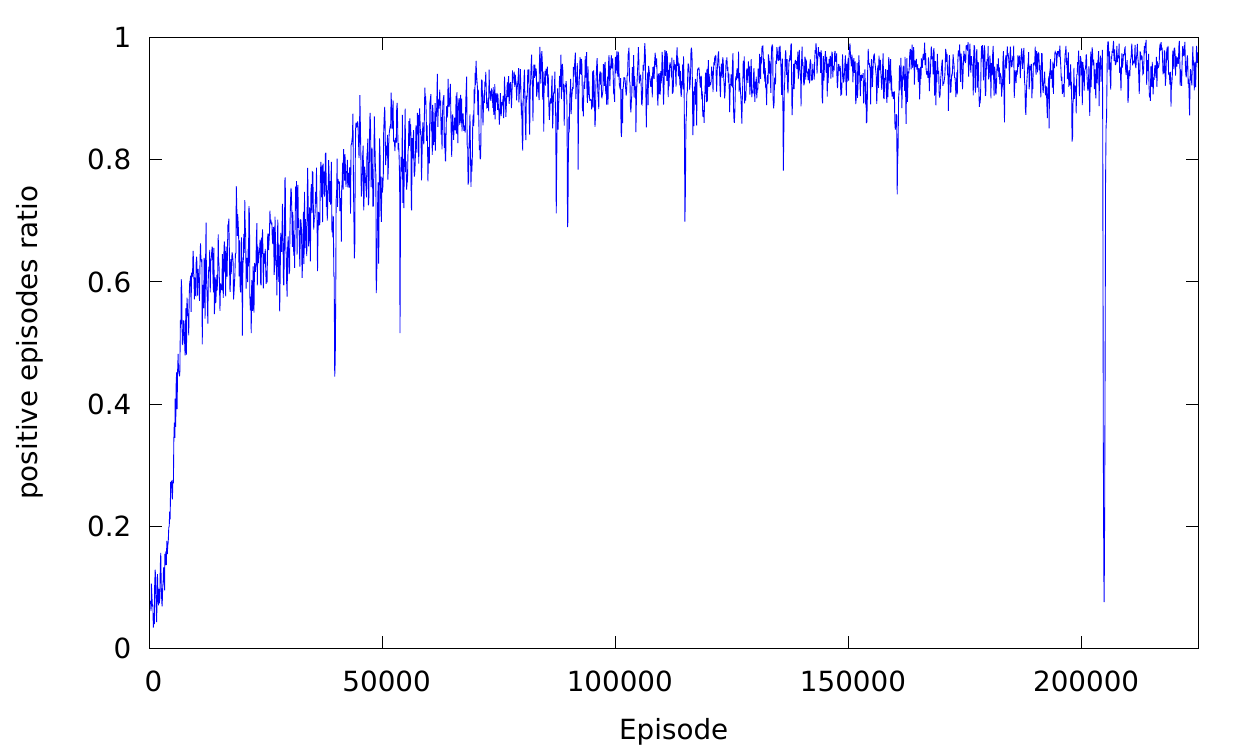}
		\label{fig:1_noar}
	}
	\subfloat[4 instances, no action repeat]{
		\includegraphics[width=0.5\columnwidth]{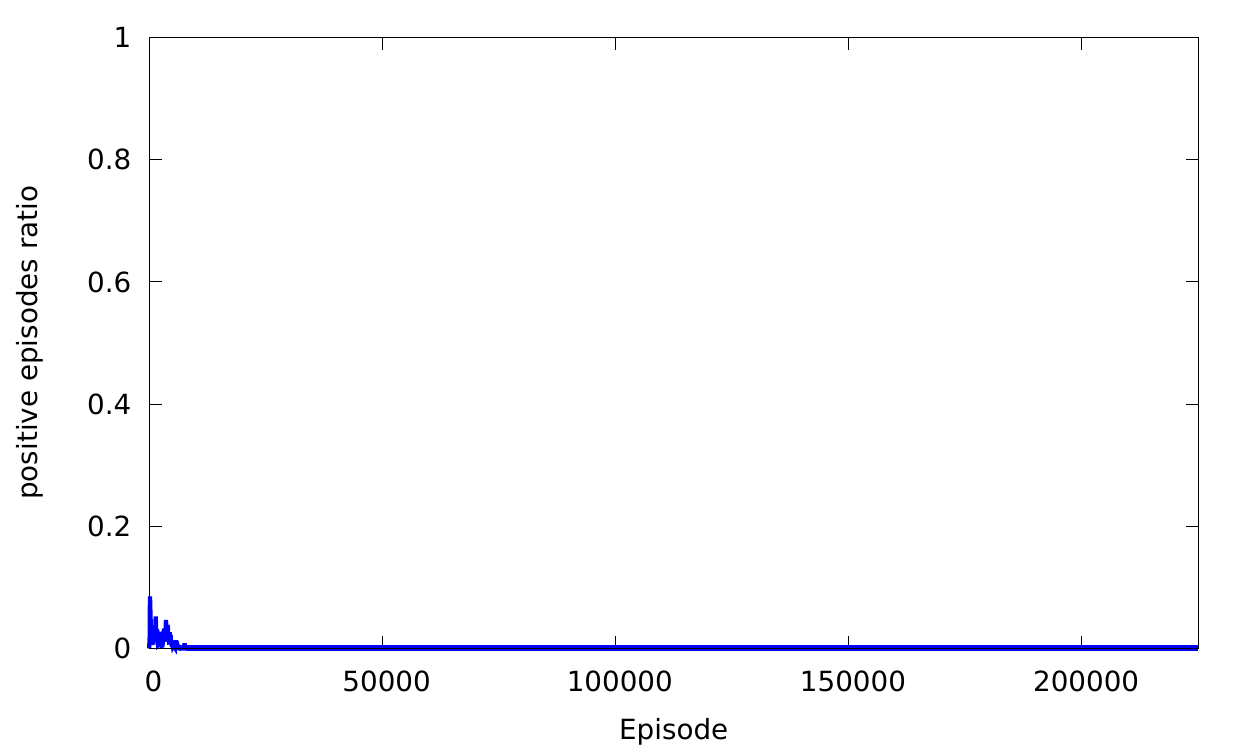}
		\label{fig:4_noar}
	}
	\caption{Learning comparison between systems trained with single or multiple environment instances, with or without action repeat. \label{fig:instances_comparison}}
\end{figure}

The learning curves, shown in Figure~\ref{fig:instances_comparison}, tell that, when using the action repeat technique, training the system with multiple instances stabilizes learning and leads to a better overall performance, even if in both cases the model converges. This confirms that reducing the correlation between updates brings a positive effect. 
It is interesting to note that in the multi-instance scenario agents start to improve their ability to reach the goal at a later stage but with a faster pace. Our explanation for this behavior is that, when running several environment instances, a higher rate of asynchronous updates coming from pseudo-random policies makes learning more problematic. Indeed, during the length of an episode, the local policy of the agent for which the updates are computed remains fixed, while the global policy receives asynchronous updates from several other agents whose behavior is still infant. However, when the policies of agents start to gain sense, their updates will be directed strongly toward a common goal due to a reduced search space.

Nonetheless, things change when action are not repeated. Indeed, while the model still converges when using a single environment instance, agents are not able to learn successfully when adopting the multi-instance setting, making the stabilizing effect of action repetition necessary.

However, the use of repeated actions brings a drawback, that is to diminish the rate at which agents interact with the environment. Indeed our empirical results, given in the graphs of Figure~\ref{fig:tests}, show that the positive episode ratio at test time increases if actions are not repeated, even if the model was trained with action repetition. In this way it is possible to take advantage of the stabilizing effect of action repetition, without loosing the chance of a finer action selection resolution.
%\vspace*{2.0em}
\section{Conclusion}
We presented an architecture able to model the longitudinal behavior of vehicles on a synthetic road scenario by using a multi-agent version of the A3C algorithm. This solution allows simulating road traffic without the need of specifying hand crafted behaviors for vehicles or collecting real data.

Agents populating the simulator sense the surrounding by a visual input made of a sequence of images representing different semantic information across the last time steps. The proposed architecture gives also the possibility to simulate the uniqueness of each real human driver by coupling a series of scalar parameters to the visual one. We exploited this feature so that aggressiveness and desired cruising speed of each driver can be regulated.  

Even if our framework is able to deal with generic junction shapes, we focused our tests on a specific scenario consisting of a single-lane roundabout with three entries: it would be interesting to see how well the system is able to generalize to different junction topologies for application in which these information are not known a priori. Moreover, it would be intriguing to extend the architecture in order to include lateral behaviors of vehicles in the simulation.

It is worth to mention, as can be seen in the previous graphs, that the proposed simulator is not collision free, but it reaches success ratios close to 1 when a safe driving is set. Even though a collision-free simulator could be desirable for some applications, the aim of this work was to simulate real driving scenarios in which errors and accidents happen: modeling them in the simulator may lead to the development of more robust high-level maneuver modules.

The time needed to train the model is not negligible and can reach several days in a desktop computer. The main issue is that  before proceeding to the subsequent time step, all the agents acting in the same environment instance need to have taken their actions: since some of them might be computing gradients, the overall performance decreases. Future works can be directed toward the design of a more efficient network architecture, as well as the development of a learning setting in which the gradient computations of all the agents in the same instance are executed simultaneously.
%\vspace*{3.0em}
\section{Acknowledgments}
We are grateful to the reviewers for their precious advices. 
Thanks also to Claudio, Roberto and Alessandra for their suggestions.

%%%%%%%%%%%%%%%%%%%%%%%%%%%%%%%%%%%%%%%%%%%%%%%%%%%%%%%%%%%%%%%%%%%%%%%%%%%%%%%%%%%%%%%%%%%%%%%%%%%%%%%%%
%% bibliography: see CFP for number of permitted pages

\bibliographystyle{reference_format}  % do not change this line!
\balance  % do not change this line -- unless you manually balance your last page
\bibliography{biblio}  % put name of your .bib file here

%%% -*-BibTeX-*-
%%% Do NOT edit. File created by BibTeX with style
%%% ACM-Reference-Format-Journals [18-Jan-2012].

\begin{thebibliography}{00}

%%% ====================================================================
%%% NOTE TO THE USER: you can override these defaults by providing
%%% customized versions of any of these macros before the \bibliography
%%% command.  Each of them MUST provide its own final punctuation,
%%% except for \shownote{}, \showDOI{}, and \showURL{}.  The latter two
%%% do not use final punctuation, in order to avoid confusing it with
%%% the Web address.
%%%
%%% To suppress output of a particular field, define its macro to expand
%%% to an empty string, or better, \unskip, like this:
%%%
%%% \newcommand{\showDOI}[1]{\unskip}   % LaTeX syntax
%%%
%%% \def \showDOI #1{\unskip}           % plain TeX syntax
%%%
%%% ====================================================================

\ifx \showCODEN    \undefined \def \showCODEN     #1{\unskip}     \fi
\ifx \showDOI      \undefined \def \showDOI       #1{#1}\fi
\ifx \showISBNx    \undefined \def \showISBNx     #1{\unskip}     \fi
\ifx \showISBNxiii \undefined \def \showISBNxiii  #1{\unskip}     \fi
\ifx \showISSN     \undefined \def \showISSN      #1{\unskip}     \fi
\ifx \showLCCN     \undefined \def \showLCCN      #1{\unskip}     \fi
\ifx \shownote     \undefined \def \shownote      #1{#1}          \fi
\ifx \showarticletitle \undefined \def \showarticletitle #1{#1}   \fi
\ifx \showURL      \undefined \def \showURL       {\relax}        \fi
% The following commands are used for tagged output and should be
% invisible to TeX
\providecommand\bibfield[2]{#2}
\providecommand\bibinfo[2]{#2}
\providecommand\natexlab[1]{#1}
\providecommand\showeprint[2][]{arXiv:#2}

\bibitem[\protect\citeauthoryear{Bansal, Krizhevsky, and Ogale}{Bansal
  et~al\mbox{.}}{2018}]%
        {waymo2018chauffeur}
\bibfield{author}{\bibinfo{person}{Mayank Bansal}, \bibinfo{person}{Alex
  Krizhevsky}, {and} \bibinfo{person}{Abhijit~S. Ogale}.}
  \bibinfo{year}{2018}\natexlab{}.
\newblock \showarticletitle{ChauffeurNet: Learning to Drive by Imitating the
  Best and Synthesizing the Worst}.
\newblock  (\bibinfo{year}{2018}).
\newblock
\showeprint[arxiv]{1812.03079}


\bibitem[\protect\citeauthoryear{Bojarski et~al\mbox{.}}{Bojarski
  et~al\mbox{.}}{2016}]%
        {bojarski2016endtoend}
\bibfield{author}{\bibinfo{person}{Mariusz Bojarski} {et~al\mbox{.}}}
  \bibinfo{year}{2016}\natexlab{}.
\newblock \showarticletitle{End to End Learning for Self-Driving Cars}.
\newblock  (\bibinfo{year}{2016}).
\newblock
\showeprint[arxiv]{1604.07316}


\bibitem[\protect\citeauthoryear{Braylan et~al\mbox{.}}{Braylan
  et~al\mbox{.}}{2015}]%
        {braylan2015frameskip}
\bibfield{author}{\bibinfo{person}{Alex Braylan} {et~al\mbox{.}}}
  \bibinfo{year}{2015}\natexlab{}.
\newblock \showarticletitle{Frame Skip Is a Powerful Parameter for Learning to
  Play Atari}. In \bibinfo{booktitle}{{\em AAAI-15 Workshop on Learning for
  General Competency in Video Games}}.
\newblock


\bibitem[\protect\citeauthoryear{Cosgun et~al\mbox{.}}{Cosgun
  et~al\mbox{.}}{2017}]%
        {cosgun2017gomentum}
\bibfield{author}{\bibinfo{person}{Akansel Cosgun} {et~al\mbox{.}}}
  \bibinfo{year}{2017}\natexlab{}.
\newblock \showarticletitle{Towards Full Automated Drive in Urban Environments:
  {A} Demonstration in GoMentum Station, California}.
\newblock   \bibinfo{volume}{abs/1705.01187} (\bibinfo{year}{2017}).
\newblock
\showeprint[arxiv]{1705.01187}


\bibitem[\protect\citeauthoryear{Dosovitskiy and Koltun}{Dosovitskiy and
  Koltun}{2017}]%
        {vladlen2017learning}
\bibfield{author}{\bibinfo{person}{Alexey Dosovitskiy} {and}
  \bibinfo{person}{Vladlen Koltun}.} \bibinfo{year}{2017}\natexlab{}.
\newblock \showarticletitle{Learning to Act by Predicting the Future}. In
  \bibinfo{booktitle}{{\em International Conference on Learning Representations
  (ICLR)}}.
\newblock


\bibitem[\protect\citeauthoryear{Fellendorf~M.}{Fellendorf~M.}{[n. d.]}]%
        {fellendorf2010vissim}
\bibfield{author}{\bibinfo{person}{Vortisch~P. Fellendorf~M.}}
  \bibinfo{year}{[n. d.]}\natexlab{}.
\newblock \showarticletitle{Microscopic Traffic Flow Simulator VISSIM}. In
  \bibinfo{booktitle}{{\em Fundamentals of Traffic Simulation}},
  \bibfield{editor}{\bibinfo{person}{Jaume Barcel{\`{o}}}} (Ed.).
\newblock


\bibitem[\protect\citeauthoryear{Goodfellow, Bengio, and Courville}{Goodfellow
  et~al\mbox{.}}{2016}]%
        {goodfellow2016dlbook}
\bibfield{author}{\bibinfo{person}{Ian Goodfellow}, \bibinfo{person}{Yoshua
  Bengio}, {and} \bibinfo{person}{Aaron Courville}.}
  \bibinfo{year}{2016}\natexlab{}.
\newblock \bibinfo{booktitle}{{\em Deep Learning}}.
\newblock \bibinfo{publisher}{MIT Press}.
\newblock
\newblock
\shownote{\url{http://www.deeplearningbook.org}.}


\bibitem[\protect\citeauthoryear{Gupta, Egorov, and Kochenderfer}{Gupta
  et~al\mbox{.}}{2017}]%
        {gupta2017cooperative}
\bibfield{author}{\bibinfo{person}{Jayesh~K. Gupta}, \bibinfo{person}{Maxim
  Egorov}, {and} \bibinfo{person}{Mykel Kochenderfer}.}
  \bibinfo{year}{2017}\natexlab{}.
\newblock \showarticletitle{Cooperative Multi-agent Control Using Deep
  Reinforcement Learning}. In \bibinfo{booktitle}{{\em Autonomous Agents and
  Multiagent Systems}}. \bibinfo{pages}{66--83}.
\newblock


\bibitem[\protect\citeauthoryear{Gupta, Vasardani, and Winter}{Gupta
  et~al\mbox{.}}{2018}]%
        {gupta2018negotiation}
\bibfield{author}{\bibinfo{person}{S. Gupta}, \bibinfo{person}{M. Vasardani},
  {and} \bibinfo{person}{S. Winter}.} \bibinfo{year}{2018}\natexlab{}.
\newblock \showarticletitle{Negotiation Between Vehicles and Pedestrians for
  the Right of Way at Intersections}.
\newblock \bibinfo{journal}{{\em IEEE Transactions on Intelligent
  Transportation Systems\/}} (\bibinfo{year}{2018}), \bibinfo{pages}{1--12}.
\newblock


\bibitem[\protect\citeauthoryear{Halevy, Norvig, and Pereira}{Halevy
  et~al\mbox{.}}{2009}]%
        {norvig2009data}
\bibfield{author}{\bibinfo{person}{Alon Halevy}, \bibinfo{person}{Peter
  Norvig}, {and} \bibinfo{person}{Fernando Pereira}.}
  \bibinfo{year}{2009}\natexlab{}.
\newblock \showarticletitle{The Unreasonable Effectiveness of Data}.
\newblock \bibinfo{journal}{{\em IEEE Intelligent Systems\/}}
  \bibinfo{volume}{24} (\bibinfo{year}{2009}), \bibinfo{pages}{8 -- 12}.
\newblock


\bibitem[\protect\citeauthoryear{Hoel, Wolff, and Laine}{Hoel
  et~al\mbox{.}}{2018}]%
        {hoel2018automated}
\bibfield{author}{\bibinfo{person}{Carl{-}Johan Hoel}, \bibinfo{person}{Krister
  Wolff}, {and} \bibinfo{person}{Leo Laine}.} \bibinfo{year}{2018}\natexlab{}.
\newblock \showarticletitle{Automated Speed and Lane Change Decision Making
  using Deep Reinforcement Learning}.
\newblock  (\bibinfo{year}{2018}).
\newblock
\showeprint[arxiv]{1803.10056}


\bibitem[\protect\citeauthoryear{Isele, Cosgun, Subramanian, and
  Fujimura}{Isele et~al\mbox{.}}{2017}]%
        {isele2017navigating}
\bibfield{author}{\bibinfo{person}{David Isele}, \bibinfo{person}{Akansel
  Cosgun}, \bibinfo{person}{Kaushik Subramanian}, {and} \bibinfo{person}{Kikuo
  Fujimura}.} \bibinfo{year}{2017}\natexlab{}.
\newblock \showarticletitle{Navigating Occluded Intersections with Autonomous
  Vehicles using Deep Reinforcement Learning}.
\newblock  (\bibinfo{year}{2017}).
\newblock
\showeprint[arxiv]{1705.01196}


\bibitem[\protect\citeauthoryear{Krajzewicz, Hertkorn, Feld, and
  Wagner}{Krajzewicz et~al\mbox{.}}{2002}]%
        {krajzewicz2002sumo}
\bibfield{author}{\bibinfo{person}{Daniel Krajzewicz}, \bibinfo{person}{Georg
  Hertkorn}, \bibinfo{person}{Christian Feld}, {and} \bibinfo{person}{Peter
  Wagner}.} \bibinfo{year}{2002}\natexlab{}.
\newblock \showarticletitle{SUMO (Simulation of Urban MObility); An open-source
  traffic simulation}. In \bibinfo{booktitle}{{\em 4th Middle East Symposium on
  Simulation and Modelling (MESM2002)}}. \bibinfo{pages}{183--187}.
\newblock


\bibitem[\protect\citeauthoryear{Li, Xu, and Zuo}{Li et~al\mbox{.}}{2015}]%
        {li2015reinforcement}
\bibfield{author}{\bibinfo{person}{X. Li}, \bibinfo{person}{X. Xu}, {and}
  \bibinfo{person}{L. Zuo}.} \bibinfo{year}{2015}\natexlab{}.
\newblock \showarticletitle{Reinforcement learning based overtaking
  decision-making for highway autonomous driving}. In \bibinfo{booktitle}{{\em
  International Conference on Intelligent Control and Information Processing
  (ICICIP)}}. \bibinfo{pages}{336--342}.
\newblock


\bibitem[\protect\citeauthoryear{Lillicrap, Hunt, Pritzel, Heess, Erez, Tassa,
  Silver, and Wierstra}{Lillicrap et~al\mbox{.}}{2015}]%
        {lillicrap2015ddpg}
\bibfield{author}{\bibinfo{person}{Timothy~P. Lillicrap},
  \bibinfo{person}{Jonathan~J. Hunt}, \bibinfo{person}{Alexander Pritzel},
  \bibinfo{person}{Nicolas Heess}, \bibinfo{person}{Tom Erez},
  \bibinfo{person}{Yuval Tassa}, \bibinfo{person}{David Silver}, {and}
  \bibinfo{person}{Daan Wierstra}.} \bibinfo{year}{2015}\natexlab{}.
\newblock \showarticletitle{Continuous control with deep reinforcement
  learning}.
\newblock  (\bibinfo{year}{2015}).
\newblock
\showeprint[arxiv]{1509.02971}


\bibitem[\protect\citeauthoryear{Liu, Hou, Mu, Yu, and Huang}{Liu
  et~al\mbox{.}}{2018}]%
        {liu2018elements}
\bibfield{author}{\bibinfo{person}{Jingchu Liu}, \bibinfo{person}{Pengfei Hou},
  \bibinfo{person}{Lisen Mu}, \bibinfo{person}{Yinan Yu}, {and}
  \bibinfo{person}{Chang Huang}.} \bibinfo{year}{2018}\natexlab{}.
\newblock \showarticletitle{Elements of Effective Deep Reinforcement Learning
  towards Tactical Driving Decision Making}.
\newblock  (\bibinfo{year}{2018}).
\newblock
\showeprint[arxiv]{1802.00332}


\bibitem[\protect\citeauthoryear{Liu, Kim, Pendleton, and Ang}{Liu
  et~al\mbox{.}}{2015}]%
        {liu2015situation}
\bibfield{author}{\bibinfo{person}{W. Liu}, \bibinfo{person}{S. Kim},
  \bibinfo{person}{S. Pendleton}, {and} \bibinfo{person}{M.~H. Ang}.}
  \bibinfo{year}{2015}\natexlab{}.
\newblock \showarticletitle{Situation-aware decision making for autonomous
  driving on urban road using online POMDP}. In \bibinfo{booktitle}{{\em IEEE
  Intelligent Vehicles Symposium (IV)}}. \bibinfo{pages}{1126--1133}.
\newblock


\bibitem[\protect\citeauthoryear{Lowe et~al\mbox{.}}{Lowe
  et~al\mbox{.}}{2017}]%
        {lowe2017multiagent}
\bibfield{author}{\bibinfo{person}{Ryan Lowe} {et~al\mbox{.}}}
  \bibinfo{year}{2017}\natexlab{}.
\newblock \showarticletitle{Multi-Agent Actor-Critic for Mixed
  Cooperative-Competitive Environments}.
\newblock  (\bibinfo{year}{2017}).
\newblock
\showeprint[arxiv]{1706.02275}


\bibitem[\protect\citeauthoryear{Mnih, Badia, Mirza, Graves, Lillicrap, Harley,
  Silver, and Kavukcuoglu}{Mnih et~al\mbox{.}}{2016}]%
        {mnih2016async}
\bibfield{author}{\bibinfo{person}{Volodymyr Mnih},
  \bibinfo{person}{Adri{\`{a}}~Puigdom{\`{e}}nech Badia},
  \bibinfo{person}{Mehdi Mirza}, \bibinfo{person}{Alex Graves},
  \bibinfo{person}{Timothy~P. Lillicrap}, \bibinfo{person}{Tim Harley},
  \bibinfo{person}{David Silver}, {and} \bibinfo{person}{Koray Kavukcuoglu}.}
  \bibinfo{year}{2016}\natexlab{}.
\newblock \showarticletitle{Asynchronous Methods for Deep Reinforcement
  Learning}. In \bibinfo{booktitle}{{\em Proceedings of the 33nd International
  Conference on Machine Learning (ICML)}}. \bibinfo{pages}{1928--1937}.
\newblock


\bibitem[\protect\citeauthoryear{Mnih, Kavukcuoglu, Silver, et~al\mbox{.}}{Mnih
  et~al\mbox{.}}{2015}]%
        {mnih2015humanlevel}
\bibfield{author}{\bibinfo{person}{Volodymyr Mnih}, \bibinfo{person}{Koray
  Kavukcuoglu}, \bibinfo{person}{David Silver}, {et~al\mbox{.}}}
  \bibinfo{year}{2015}\natexlab{}.
\newblock \showarticletitle{Human-level control through deep reinforcement
  learning}.
\newblock \bibinfo{journal}{{\em Nature\/}} \bibinfo{volume}{518},
  \bibinfo{number}{7540} (\bibinfo{date}{Feb.} \bibinfo{year}{2015}),
  \bibinfo{pages}{529--533}.
\newblock


\bibitem[\protect\citeauthoryear{Packard and Worth}{Packard and Worth}{2018}]%
        {cairographics}
\bibfield{author}{\bibinfo{person}{Keith Packard} {and} \bibinfo{person}{Carl
  Worth}.} \bibinfo{year}{2003--2018}\natexlab{}.
\newblock \bibinfo{title}{Cairo graphics library}.
\newblock \bibinfo{howpublished}{\url{https://www.cairographics.org/}}.
  (\bibinfo{year}{2003--2018}).
\newblock


\bibitem[\protect\citeauthoryear{Paxton, Raman, Hager, and Kobilarov}{Paxton
  et~al\mbox{.}}{2017}]%
        {paxton2017combining}
\bibfield{author}{\bibinfo{person}{Chris Paxton}, \bibinfo{person}{Vasumathi
  Raman}, \bibinfo{person}{{Gregory D.} Hager}, {and} \bibinfo{person}{Marin
  Kobilarov}.} \bibinfo{year}{2017}\natexlab{}.
\newblock \showarticletitle{Combining neural networks and tree search for task
  and motion planning in challenging environments}. In \bibinfo{booktitle}{{\em
  IEEE/RSJ International Conference on Intelligent Robots and Systems (IROS)}},
  Vol.~\bibinfo{volume}{2017-September}. \bibinfo{pages}{6059--6066}.
\newblock


\bibitem[\protect\citeauthoryear{Schulman, Levine, Moritz, Jordan, and
  Abbeel}{Schulman et~al\mbox{.}}{2015}]%
        {schulman2015trpo}
\bibfield{author}{\bibinfo{person}{John Schulman}, \bibinfo{person}{Sergey
  Levine}, \bibinfo{person}{Philipp Moritz}, \bibinfo{person}{Michael~I.
  Jordan}, {and} \bibinfo{person}{Pieter Abbeel}.}
  \bibinfo{year}{2015}\natexlab{}.
\newblock \showarticletitle{Trust Region Policy Optimization}. In
  \bibinfo{booktitle}{{\em International Conference on Machine Learning
  (ICML)}}.
\newblock


\bibitem[\protect\citeauthoryear{Shalev{-}Shwartz, Shammah, and
  Shashua}{Shalev{-}Shwartz et~al\mbox{.}}{2016}]%
        {shwartz2016safe}
\bibfield{author}{\bibinfo{person}{Shai Shalev{-}Shwartz},
  \bibinfo{person}{Shaked Shammah}, {and} \bibinfo{person}{Amnon Shashua}.}
  \bibinfo{year}{2016}\natexlab{}.
\newblock \showarticletitle{Safe, Multi-Agent, Reinforcement Learning for
  Autonomous Driving}.
\newblock  (\bibinfo{year}{2016}).
\newblock
\showeprint[arxiv]{1610.03295}


\bibitem[\protect\citeauthoryear{Sutton and Barto}{Sutton and Barto}{2018}]%
        {suttonbarto2018rlbook}
\bibfield{author}{\bibinfo{person}{Richard~S. Sutton} {and}
  \bibinfo{person}{Andrew~G. Barto}.} \bibinfo{year}{2018}\natexlab{}.
\newblock \bibinfo{booktitle}{{\em Reinforcement Learning: An Introduction}}.
\newblock \bibinfo{publisher}{MIT Press}.
\newblock


\bibitem[\protect\citeauthoryear{Tampuu et~al\mbox{.}}{Tampuu
  et~al\mbox{.}}{2015}]%
        {tampuu2015multiagent}
\bibfield{author}{\bibinfo{person}{Ardi Tampuu} {et~al\mbox{.}}}
  \bibinfo{year}{2015}\natexlab{}.
\newblock \showarticletitle{Multiagent Cooperation and Competition with Deep
  Reinforcement Learning}.
\newblock  (\bibinfo{year}{2015}).
\newblock
\showeprint[arxiv]{1511.08779}


\bibitem[\protect\citeauthoryear{Treiber, Hennecke, and Helbing}{Treiber
  et~al\mbox{.}}{2000}]%
        {treiber2000idm}
\bibfield{author}{\bibinfo{person}{M Treiber}, \bibinfo{person}{A Hennecke},
  {and} \bibinfo{person}{D Helbing}.} \bibinfo{year}{2000}\natexlab{}.
\newblock \showarticletitle{Congested Traffic States in Empirical Observations
  and Microscopic Simulations}.
\newblock \bibinfo{journal}{{\em Phys. Rev. E\/}}  \bibinfo{volume}{62}
  (\bibinfo{year}{2000}), \bibinfo{pages}{1805--1824}.
\newblock


\bibitem[\protect\citeauthoryear{Urmson et~al\mbox{.}}{Urmson
  et~al\mbox{.}}{2008}]%
        {urmson2008boss}
\bibfield{author}{\bibinfo{person}{Chris Urmson} {et~al\mbox{.}}}
  \bibinfo{year}{2008}\natexlab{}.
\newblock \showarticletitle{Autonomous Driving in Urban Environments: Boss and
  the Urban Challenge}.
\newblock \bibinfo{journal}{{\em J. Field Robot.\/}} \bibinfo{volume}{25},
  \bibinfo{number}{8} (\bibinfo{date}{Aug.} \bibinfo{year}{2008}),
  \bibinfo{pages}{425--466}.
\newblock


\bibitem[\protect\citeauthoryear{van~der Horst and Hogema}{van~der Horst and
  Hogema}{1994}]%
        {vanderhorst1994ttc}
\bibfield{author}{\bibinfo{person}{Richard van~der Horst} {and}
  \bibinfo{person}{Jeroen Hogema}.} \bibinfo{year}{1994}\natexlab{}.
\newblock \showarticletitle{Time-To-Collision and Collision Avoidance Systems}.
  In \bibinfo{booktitle}{{\em 6th ICTCT workshop}}.
\newblock


\bibitem[\protect\citeauthoryear{Watkins}{Watkins}{1989}]%
        {watkins1989qlearning}
\bibfield{author}{\bibinfo{person}{Christopher John Cornish~Hellaby Watkins}.}
  \bibinfo{year}{1989}\natexlab{}.
\newblock {\em \bibinfo{title}{Learning from Delayed Rewards}}.
\newblock \bibinfo{thesistype}{Ph.D. Dissertation}. \bibinfo{school}{King's
  College}, \bibinfo{address}{Cambridge, UK}.
\newblock


\bibitem[\protect\citeauthoryear{Williams}{Williams}{1992}]%
        {williams1992reinforce}
\bibfield{author}{\bibinfo{person}{Ronald~J. Williams}.}
  \bibinfo{year}{1992}\natexlab{}.
\newblock \showarticletitle{Simple Statistical Gradient-Following Algorithms
  for Connectionist Reinforcement Learning}.
\newblock \bibinfo{journal}{{\em Mach. Learn.\/}} \bibinfo{volume}{8},
  \bibinfo{number}{3-4} (\bibinfo{date}{May} \bibinfo{year}{1992}),
  \bibinfo{pages}{229--256}.
\newblock


\end{thebibliography}

\end{document}